\documentclass[pra,twocolumn,amsmath,amssymb,superscriptaddress]{revtex4-1}
\usepackage{epsfig,amsmath}
\usepackage{subfigure}
\usepackage{graphicx}
\usepackage{dcolumn}
\usepackage{stmaryrd}
\usepackage{mathrsfs}
\usepackage{pifont}
\usepackage{amsthm}
\usepackage{amssymb}
\usepackage{bm}
\usepackage{latexsym}
\usepackage[colorlinks=true,linkcolor=blue,citecolor=blue]{hyperref}
\usepackage{color}
\usepackage{epstopdf}
\usepackage{booktabs}
\usepackage{amsmath}
\usepackage{threeparttable}

\begin{document}

\title{Stabilizing a single-magnon state by optimizing magnon blockade}

\author{Zhu-yao Jin}
\affiliation{School of Physics, Zhejiang University, Hangzhou 310027, Zhejiang, China}

\author{Jun Jing}
\email{Email address: jingjun@zju.edu.cn}
\affiliation{School of Physics, Zhejiang University, Hangzhou 310027, Zhejiang, China}

\date{\today}

\begin{abstract}
A stable and high-quality single-magnon state is desired by the single-magnon source for quantum information application with a macroscopic spin system. We consider a hybrid system where a magnon mode is directly coupled to a nonresonant superconducting qubit via the exchange interaction. The magnon and qubit are under the driving and probing fields with the same frequency, respectively. We find that the single-magnon probability $P_1$ can be maximized when the product of the magnon-driving field detuning and the qubit-probing field detuning is equivalent to the square of the magnon-qubit coupling strength, $\Delta_q\Delta_m=J^2$. Then, the double-magnon probability $P_2$ can be minimized by tuning the ratio of the probing intensity to the driving intensity and the relative phase between the two fields. Under these optimized conditions with accessible strong driving intensity and low decay rate, strong magnon blockade gives rise to a stable single-magnon state with a high quality. It features a large brightness (the single-magnon probability) $P_1\approx0.40$ and a high purity (the equal-time second-order correlation function) $g^{(2)}(0)\sim10^{-5}$. The two indicators as a whole prevail over the existing results for photon, phonon, and magnon modes with respect to a stable single-quantum state. The optimized conditions with a scalable modification $\Delta_q\Delta_m\approx NJ^2$ apply to the situation when one focus on only one of the $N$ magnon modes that are simultaneously coupled to a common qubit.
\end{abstract}

\maketitle

\section{Introduction}

As a quantized spin wave, magnon can be generated by the collective excitation of a large number of spins in the yttrium iron garnet (YIG) sphere~\cite{Serga2010YIG,AV2015nature,zheng2023tutorial}, a ferrimagnetic insulator with an extremely high spin density and record low damping rate. Recently, the YIG sphere~\cite{Serga2010YIG} becomes an essential component of various hybrid magnonic systems that have attracted considerable attentions in advancing quantum information processing~\cite{Wallquist2009Hybrid,Kimble2008internet} and quantum technology~\cite{Gershon2015Quantum}. At least two reasons support that the hybrid magnonic systems are more promising than the optomechanical systems: (1) by virtue of the great tunability in frequency~\cite{Soykal2010Strong,Huebl2013High,Zhang2014Strongly,Tabuchi2014Hybridizing,AV2015nature}, magnons can be integrated with plentiful quantum elements, including phonon modes~\cite{Zhang2016Cavity,Li2018magnon,Shen2022Coherent}, microwave photons~\cite{Soykal2010Strong,Huebl2013High,Goryachev2014High,Zhang2014Strongly,Tabuchi2014Hybridizing,
Mergenthaler2017Strong,Bourhill2016Ultrahigh,Kostylev2015Superstrong}, optical photons~\cite{Osada2016Cavity,Zhang2016Optomagnonics,Haigh2016Triple,Osada2018Brillouin,Parvini2020Antiferromagnetic}, and superconducting qubits~\cite{Yutaka2015Coherent,Dany2017Resolving,Dany2020Entanglement,Xu2023Quantum,Wolski2020Dissipation,
Kounalakis2022Analog}, by various coupling schemes such as pressure-like interaction and magnetic dipole interaction~\cite{Imamoifmmode2009Cavity}; (2) owing to the high spin density of YIG, strong~\cite{Huebl2013High,Zhang2014Strongly,Tabuchi2014Hybridizing,Mergenthaler2017Strong} or even ultrastrong interaction~\cite{Goryachev2014High,Bourhill2016Ultrahigh,Kostylev2015Superstrong} between a magnon and a microwave cavity, and strong indirect~\cite{Yutaka2015Coherent,Dany2017Resolving,Dany2020Entanglement,Wolski2020Dissipation,Xu2023Quantum} (mediated by a microwave cavity) or strong direct interactions~\cite{Kounalakis2022Analog} between a magnon and a superconducting qubit have been established. One can thus exploit magnon, the macroscopic spin system, to carry out diversified tasks of information processing in both classical and quantum regimes.

A lot of exotic properties of the hybrid magnonic systems have been detected in the classical regime, such as bistability~\cite{Wang2018bistability,Zhang2019Theory}, multistability~\cite{Shen2021Long,Shen2021Long}, magnonic frequency comb~\cite{Wang2021Magnonic,Rao2023Unveiling,Wang2022Twisted}, and magnonic Penrose superradiance~\cite{Wang2022Twisted}. In the quantum regime, many proposals were put forward to generate entangled states, such as entangling the magnon-photon-phonon system~\cite{Li2018magnon,Amazioug2023enhancement}, generating the Bell states between a magnon and photon~\cite{Yuan2020Steady,Qi2022Generation}, and entangling two magnons through a microwave cavity~\cite{Yuan2020Enhancement,Azimi2021Magnon,Ren2022Long}. Other applications include quantum transducer~\cite{Hisatomis2016Bidirectional,Zhu2020Waveguide} and quantum sensing~\cite{Wolski2020Dissipation,Cai2021Microwave,Zhang2023Detection}. The quantum control over the macroscopic spin system~\cite{Dany2020Entanglement,Xu2023Quantum} is fundamentally based on the operation at the level of a single magnon, similar to that over the superconducting resonator~\cite{Hofheinz2008Generation}, the optomechanical resonator~\cite{Connell2010Quantum,Sungkun2017Hanbury}, and the acoustic-wave systems~\cite{Chu2018Creation}. From the ground state, the magnon Fock states and their superposition~\cite{Bittencourt2019Magnon,Sun2021Remote,Sharma2022protocol,He2023Generation} can be nondeterministically heralded or temporarily generated through direct or indirect exchange interaction.

In the long-time limit, a stable single-magnon state could be created by magnon blockade~\cite{Liu2019Magnon,Xie2020Quantum,Wu2021Phase,Wang2022Hybrid,Yuan2020Magnon}, working as a solid single-quantum source. Similar to the single-photon source~\cite{Lounis2005Single}, by which a single photon can be emitted at the moments desired by the user~\cite{Eisamann2011Invited}, the steady single-magnon state relies on optimized blockade conditions and is irrespective of the initial state. Enlightened by the statistics in a photon or phonon blockade~\cite{Rabl2011Photon,Lounis2005Single,Milburn2015Quantum,Kar2013Single,Tang2015Quantum,Flayac2017Unconventional,
Tang2019Strong,Li2021Strong,Couteau2023Applications,Shijders2018Observation,Wang2020Photon,Rabl2011Photon}, the steady single-magnon state is characterized with a significant occupation on the single-quantum state $P_1$ and a low equal-time second-order correlation function $g^{(2)}(0)$. The standard criterion for confirming a steady single-magnon state is $g^{(2)}(0)<0.5$~\cite{Couteau2023Applications}, since $g^{(2)}(0)=0.5$ when the magnon is at the Fock state $|2\rangle$. A perfect Fock state $|1\rangle$~\cite{Couteau2023Applications} implies that $P_1\rightarrow1$ and $g^{(2)}(0)\rightarrow0$. However, most existing proposals focus only on optimizing the $g^{(2)}$ function, and the magnitude of $P_1$ is typically in the range of $(0.01, 0.1)$.

In this paper, we target to generate a stable and high-quality single-magnon state by optimizing the magnon blockade in a system that the magnon is coupled to a nonresonant superconducting qubit via the exchange interaction~\cite{Kounalakis2022Analog}. In particular, we seek to amplify the single-magnon probability $P_1$ and suppress the double-magnon probability $P_2$ when the system is pushed to the strong blockade regime. The optimized conditions are confirmed to be associated with the qubit-probing field detuning, the magnon-driving field detuning, the magnon-qubit coupling strength, the field intensity ratio, and the relative phase between the two fields. Using the fact that $P_1$ is roughly proportional to the driving-field intensity and inversely proportional to the system decay rate, one can achieve $P_1\sim0.40$ and $g^{(2)}(0)\sim10^{-5}$ with the achievable parameters. In regard of the two indicators, the best and balanced results for a steady single-photon state are found to be $P_1\sim0.10$ and $g^{(2)}(0)\sim10^{-2}$, which were constructed by a two-photon absorption process of the environment~\cite{Zhou2022Environmentally}. Straightforwardly, our proposal can be generalized to a model of $N$ magnons coupled to a common qubit~\cite{Kong2021Magnon,Kong2022Remote,Li2022Tunable}, in which every magnon mode can be stabilized as a high-quality single-excitation resource under the scaled condition for maximizing $P_1$.

The rest of this paper is structured as follows. In Sec.~\ref{modelandHam}, we introduce the hybrid magnon-qubit system under the external driving and probing fields. In Sec.~\ref{SingleMagnonState}, we use an approximate model with a non-Hermitian Hamiltonian to analytically derive the optimal conditions for the strong magnon blockade, and they are numerically confirmed in Sec.~\ref{verify}. Next, in Sec.~\ref{highquality}, we approach the high-quality and stable single-magnon state through numerical optimization. In Sec.~\ref{generalmagnon}, we discuss the general case with $N$ magnons. The whole work is concluded in Sec.~\ref{Conclusion}.

\section{Model and Hamiltonian}\label{modelandHam}

\begin{figure}[htbp]
\centering
\includegraphics[width=0.9\linewidth]{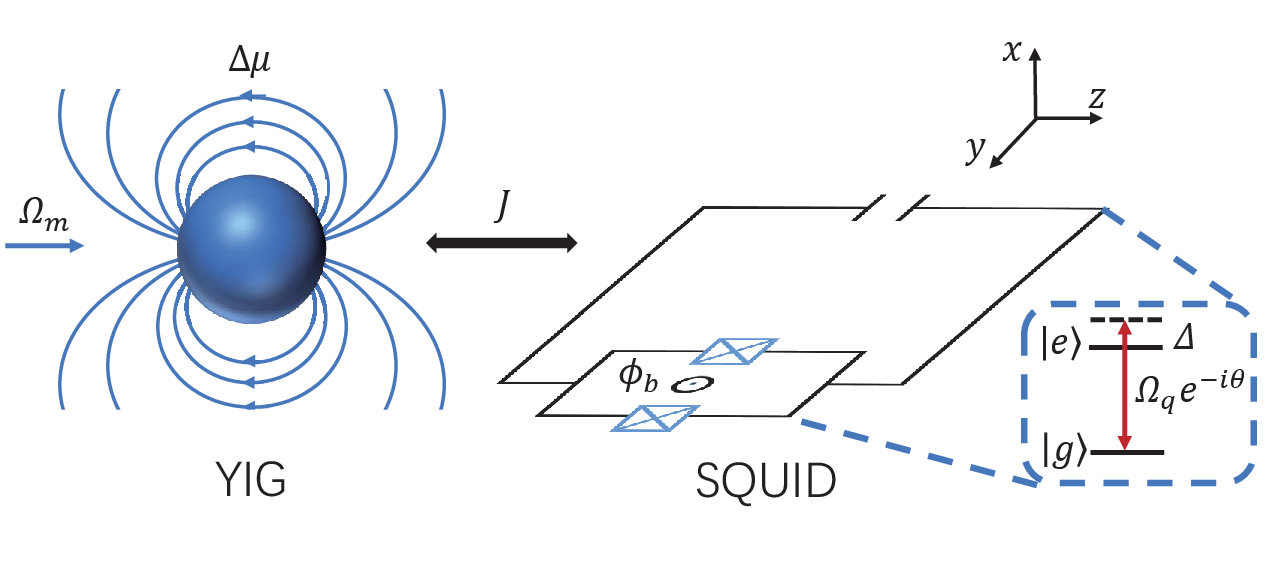}
\caption{Sketch of the hybrid magnon-qubit system, in which the ferromagnetic YIG sphere is directly coupled to a superconducting transmon qubit~\cite{Kounalakis2022Analog}. The magnon and the qubit are, respectively, under the driving field with Rabi frequency $\Omega_m$ and the probing field with Rabi frequency $\Omega_q$. The two fields are out of phase with $\theta$. }\label{model1}
\end{figure}

We consider a hybrid magnon-qubit system~\cite{Kounalakis2022Analog} as shown in Fig.~\ref{model1}, where a YIG sphere is directly coupled to a superconducting transmon qubit via the transversal (exchange) interaction $J$. The original Hamiltonian reads ($\hbar\equiv1$)
\begin{equation}\label{Ham}
H=\omega_q\sigma_+\sigma_-+\omega_mm^\dagger m+J(m\sigma_++m^\dagger\sigma_-).
\end{equation}
The qubit is formed by a superconducting quantum interference device (SQUID) loop and a capacitor in parallel. And, the SQUID loop is interrupted by two Josephson junctions. The flux through the SQUID loop consists of the external flux $\Phi_b$ induced by the control lines carrying direct and alternating currents and the flux $\Phi(\Delta\mu)$ caused by the magnetic fluctuation $\Delta\mu$. The latter establishes the direct magnon-qubit interaction, including the transversal and longitudinal parts. The transversal part follows the exchange interaction form~\cite{Yutaka2015Coherent,Jaynes1963comparison}, which plays a key role in our proposal. And the longitudinal interaction is analogous to the radiation pressure in optomechanics~\cite{Shevchuk2017Strong,Rodrigues2019coupling,Kounalakis2020Flux}. Both of them can be tuned through modulating the external flux $\Phi_b$ and the magnon-qubit distance $d$. When the external flux $\Phi_b/\Phi_0\approx0.5$ ($\Phi_0$ is the magnetic flux quantum)~\cite{Kounalakis2022Analog}, the longitudinal interaction can be switched off. In this situation, the transversal interaction can be tuned up to about $J/2\pi\sim35$ MHz~\cite{Kounalakis2022Analog} by properly positioning the magnon and qubit. As for the indirect coupling mediated by the cavity~\cite{Yutaka2015Coherent,Dany2017Resolving,Dany2020Entanglement,Xu2023Quantum}, the magnon-qubit exchange interaction is about $J/2\pi\sim15$ MHz. The qubit frequency $\omega_q/2\pi$ is tunable in the range from $1$ to $10$ GHz through modulating the external flux $\Phi_b$ and the SQUID asymmetry $\alpha$~\cite{Kounalakis2022Analog}, e.g., $\omega_q/2\pi\sim1.5$ GHz for $\Phi_b/\Phi_0\approx0.5$ and $\alpha=0.01$. In comparison to its optomechanical counterpart~\cite{Rabl2011Photon,Xie2017Phonon}, the magnon frequency $\omega_m=\gamma B$ with $\gamma$ the gyromagnetic ratio is flexibly tunable in the range of $1-10$ GHz through the external bias magnetic field $B$. In terms of the Gilbert damping constant $\alpha_G\sim10^{-5}-10^{-4}$~\cite{Kounalakis2022Analog}, the damping rate of magnon $\kappa_m/2\pi\sim\omega_m\alpha_G$ is in the order of $0.1-1$ MHz~\cite{Tabuchi2014Hybridizing,Kounalakis2022Analog}.

To push the magnon mode into the single-quantum state through magnon blockade, the qubit and magnon are respectively driven by a probing field and a driving field, which are assumed to be resonant in frequency $\omega$. Typically, $\omega$ is on the order of gigahertz~\cite{Wang2018bistability}. We consider a relative phase $\theta$ between the probing and driving fields. Then, the total Hamiltonian becomes
\begin{equation}\label{Hamtot}
\begin{aligned}
H_{\rm tot}&=\omega_q\sigma_+\sigma_-+\omega_mm^\dagger m+J(m\sigma_++m^\dagger\sigma_-)\\
&+\left[\Omega_mm^\dagger e^{-i\omega t}+\Omega_q\sigma_+e^{-i(\omega t+\theta)}+{\rm H.c.}\right],
\end{aligned}
\end{equation}
where the probing intensity $\Omega_q=k\sqrt{P_d}$ with $k=103$ MHz/mW$^{1/2}$~\cite{Wang2019simulation} and the field power $P_d$ as strong as $350$ mW~\cite{Wang2018bistability}, and the driving intensity $\Omega_m=\sqrt{2S}\Omega_s$ with $S$ the total spin number of the macroscopic spin system and $\Omega_s$ the coupling strength of the driving field with the macrospin~\cite{Wang2016magnon}. In the rotating frame with respect to $H'=\omega(m^\dagger m+\sigma_+\sigma_-)$, the effective Hamiltonian reads
\begin{equation}\label{Ham_eff}
\begin{aligned}
H_{\rm eff}&=\Delta_q\sigma_+\sigma_-+\Delta_mm^\dagger m+J(m\sigma_++m^\dagger\sigma_-)\\
&+\Omega_m(m^{\dagger}+m)+\Omega_q(\sigma_+e^{-i\theta}+\sigma_-e^{i\theta}),
\end{aligned}
\end{equation}
where $\Delta_q\equiv\omega_q-\omega$ and $\Delta_m\equiv\omega_m-\omega$. Here, we consider a more general situation than the existing proposals~\cite{Liu2019Magnon,Li2021Strong,Jin2023magnon,Tang2015Quantum} where the magnon and qubit are nonresonant, i.e., $\Delta_q\neq\Delta_m$.

\section{Optimal conditions for stabilizing single-magnon state}

\subsection{Maximize single-magnon probability}\label{SingleMagnonState}

Statistically, the single-magnon state could be characterized with brightness and purity. The former is defined as the average occupation
\begin{equation}\label{bright}
\bar{n}=\langle m^\dagger m\rangle=\sum_{n>0}nP_n
\end{equation}
with the probability in the magnon-number state
\begin{equation}\label{Pdefine}
P_n\equiv\langle n|\rho_m|n\rangle,
\end{equation}
and the latter is suggested by the equal-time second-order correlation function~\cite{Carmichael1999statistical,Eleuch2008Photon}
\begin{equation}\label{g2defineGeneral}
g^{(2)}(0)=\frac{\langle m^\dagger m^\dagger mm\rangle}{\langle m^\dagger m\rangle^2}.
\end{equation}
Here, the expectation values are evaluated by the reduced density matrix for the magnon mode $\rho_m$, which can be obtained by partially tracing the full density matrix $\rho$ of the steady state over the degree of freedom of the qubit.

When the magnon is pushed to the blockade regime with a strong antibunching effect, $g^{(2)}(0)\rightarrow0$, i.e., $P_0, P_1\gg\sum_nP_{n>1}$ and $P_2\gg P_{n>2}$, the brightness of the single-magnon state becomes
\begin{equation}\label{brightdefine}
\bar{n}\approx P_1,
\end{equation}
and the purity can be expressed by
\begin{equation}\label{g2_P}
g^{(2)}(0)=\frac{\sum_{n=2}^{\infty}n(n-1)P_n}{\sum_{n=1}^{\infty}(nP_n)^2}\approx\frac{2P_2}{P_1^2}.
\end{equation}
In the strong-coupling limit, i.e., $J\gg\Omega_m,\Omega_q$, an analytical model~\cite{Kar2013Single,Tang2015Quantum,
Flayac2017Unconventional,Tang2019Strong,Liu2019Magnon,Xie2020Quantum,Wu2021Phase,Wang2022Hybrid} can be used to estimate the wave function of the composite system and then to derive analytically the optimal conditions for amplifying $P_1$ and suppressing $P_2$. Despite the fact that this model cannot be fully correct in the presence of dissipation, the resulting optimal conditions for $P_1$ and $P_2$ can be confirmed with the master-equation approach~\cite{Kar2013Single}. Specifically, the system dynamics can be approximately determined by the non-Hermitian Hamiltonian
\begin{equation}\label{Hamnon1}
H_{\rm non}=H_{\rm eff}-i\frac{\kappa}{2}(\sigma_+\sigma_-+m^\dagger m)
\end{equation}
with the effective Hamiltonian $H_{\rm eff}$ in Eq.~(\ref{Ham_eff}). And, the composite system could be truncated to a subspace with a few excitations (not greater than two), i.e., the system evolution can be approximately described by
\begin{equation}\label{purestate1}
|\psi\rangle=C_{g0}|g0\rangle+C_{g1}|g1\rangle+C_{e0}|e0\rangle+C_{e1}|e1\rangle+C_{g2}|g2\rangle,
\end{equation}
where $C_\alpha$ ($\alpha=g0, g1, e0, e1, g2$) is the probability amplitude and $|g(e)\rangle$ and $|n\rangle$ denote the ground (excited) state of qubit and the Fock state of the magnon mode, respectively. Then, the relevant dynamical equations can be written as
\begin{equation}\label{SchorEq1}
\begin{aligned}
&i\dot{C}_{e0}=JC_{g1}+\Omega_qe^{-i\theta}C_{g0}+\tilde{\Delta}_qC_{e0},\\
&i\dot{C}_{g1}=JC_{e0}+\Omega_mC_{g0}+\tilde{\Delta}_mC_{g1},\\
&i\dot{C}_{e1}=\sqrt{2}JC_{g2}+\Omega_qe^{-i\theta}C_{g1}+\Omega_mC_{e0}\\
&+(\tilde{\Delta}_q+\tilde{\Delta}_m)C_{e1},\\
&i\dot{C}_{g2}=\sqrt{2}JC_{e1}+\sqrt{2}\Omega_mC_{g1}+2\tilde{\Delta}_mC_{g2},
\end{aligned}
\end{equation}
where $\tilde{\Delta}_{q, m}\equiv\Delta_{q, m}-i\kappa/2$. The steady-state solutions yield
\begin{subequations}
\begin{align}\label{SteadyAmpa}
C_{g1}&=\frac{J\Omega_qe^{-i\theta}-\Omega_m\tilde{\Delta}_q}
{\tilde{\Delta}_q\tilde{\Delta}_m-J^2},\\ \label{SteadyAmpb}
C_{g2}&=\frac{\sqrt{2}(AC_{g1}-B)}{2\tilde{\Delta}_m(\tilde{\Delta}_q
+\tilde{\Delta}_m)-2J^2}\\ \label{SteadyAmpc}
C_{e1}&=\frac{-\sqrt{2}J\tilde{\Delta}_qC_{g2}-(\Omega_q\tilde{\Delta}_qe^{-i\theta}-\Omega_mJ)C_{g1}}
{\tilde{\Delta}_q(\tilde{\Delta}_q+\tilde{\Delta}_m)}\\
&+\frac{B}{J(\tilde{\Delta}_q+\tilde{\Delta}_m)}\nonumber
\end{align}
\end{subequations}
with the coefficients $A$ and $B$
\begin{equation}\label{AB}
\begin{aligned}
A&=J\Omega_qe^{-i\theta}-\left(\tilde{\Delta}_q+\tilde{\Delta}_m+\frac{J^2}{\tilde{\Delta}_q}\right)\Omega_m,\\
B&=\frac{J\Omega_m\Omega_qe^{-i\theta}}{\tilde{\Delta}_q}.
\end{aligned}
\end{equation}
The single-magnon and double-magnon probabilities can thus be expressed as
\begin{equation}\label{P1P2single}
\begin{aligned}
P_1&=|C_{g1}|^2+|C_{e1}|^2\approx|C_{g1}|^2,\\
P_2&=|C_{g2}|^2
\end{aligned}
\end{equation}
under the approximation $|C_{g1}|\gg|C_{e1}|$. In the weak decay regime, i.e., $\kappa\ll\Delta_{m, q}, J$, the population on the state $|g1\rangle$ can be maximized by
\begin{equation}\label{OptDelta1}
\Delta_q\Delta_m\approx\tilde{\Delta}_q\tilde{\Delta}_m=J^2
\end{equation}
due to Eq.~(\ref{SteadyAmpa}). Then, the amplitude magnitude $|C_{g2}|$ in Eq.~(\ref{SteadyAmpb}) can be optimized by minimizing the coefficient $A$ for $C_{g1}$. With Eq.~(\ref{AB}), Eq.~(\ref{OptDelta1}), and $\theta\ll 1$, the condition reads
\begin{equation}\label{OptOq}
\frac{\Omega_q}{\Omega_m}\approx\frac{\Delta_q+2\Delta_m}{J}.
\end{equation}
Consequently, we have
\begin{equation}\label{Cg2theta}
C_{g2}\approx\frac{i\sqrt{2}\left[4\Delta_m(\Delta_q+2\Delta_m)\theta-(\Delta_q+7\Delta_m)\kappa\right]\Omega_m^2}
{2\Delta_m^2(\Delta_q+\Delta_m)\kappa},
\end{equation}
by which $|C_{g2}|$ can be further reduced if
\begin{equation}\label{OptTheta}
\theta\rightarrow\theta_{\rm opt}=\frac{(\Delta_q+7\Delta_m)\kappa}{4\Delta_m(\Delta_q+2\Delta_m)}.
\end{equation}

\subsection{Numerical verification of optimal conditions}\label{verify}

We here numerically verify the optimal conditions presented in Eqs.~(\ref{OptDelta1}), (\ref{OptOq}), and (\ref{OptTheta}) for the strong magnon blockade. The first condition primarily targets increasing the single-magnon occupation $P_1$, and the second and third one contribute to decreasing the double-magnon occupation $P_2$ as well as the equal-time second-order correlation function $g^{(2)}{(0)}$.

The decoherence dynamics towards the steady state can be calculated by the master equation~\cite{Carmichael1999statistical},
\begin{equation}\label{masterequationGeneral}
\frac{\partial}{\partial t}\rho=-i[H_{\rm eff}, \rho]+\frac{\kappa_m}{2}\mathcal{L}_m[\rho]+\frac{\kappa_q}{2}\mathcal{L}_{\sigma_-}[\rho],
\end{equation}
where $H_{\rm eff}$ is the effective Hamiltonian in Eq.~(\ref{Ham_eff}) and the dissipator is defined as the Lindblad superoperators $\mathcal{L}_o[\rho]=2o\rho o^\dagger-o^\dagger o\rho-\rho o^\dagger o$~\cite{Scully1997quantum} with $o=m, \sigma_-$ indicating, respectively, the decay channels for the magnon mode and the qubit. In our simulation, the Hilbert space is truncated up to a limited number of Fock state $|N\rangle$ with $N=10$. The decay rates of the magnon and qubit are set as $\kappa_m=\kappa_q=\kappa$ for simplicity.

\begin{figure}[htbp]
\centering
\includegraphics[width=0.9\linewidth]{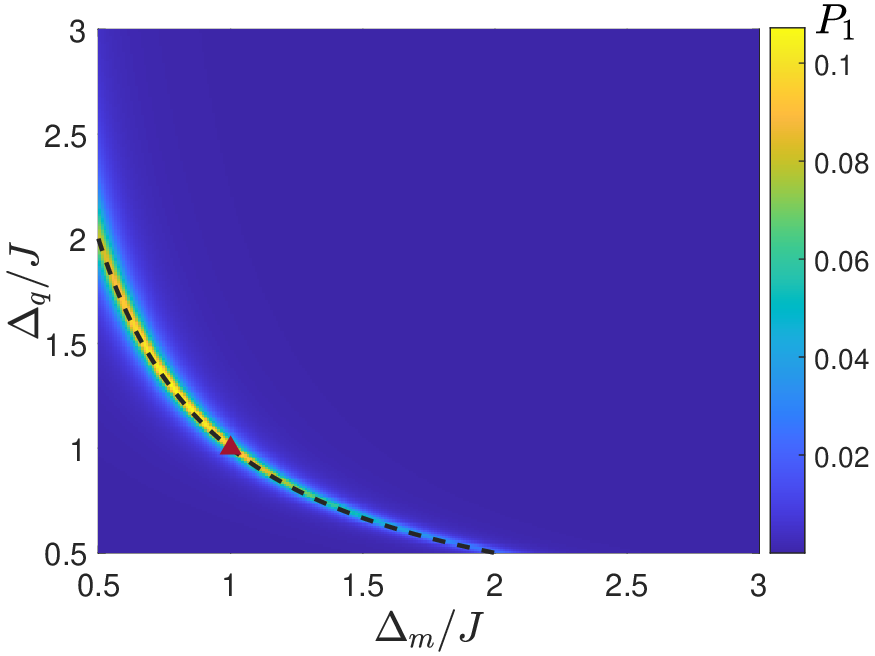}
\caption{Single-magnon probability $P_1$ vs the ratio of the detuning $\Delta_q$ to the coupling strength $J$ and the ratio of the detuning $\Delta_m$ to the coupling strength $J$. The black dashed line follows the optimal condition about $\Delta_q\Delta_m$ in Eq.~(\ref{OptDelta1}). The red triangle describes the resonant case $\Delta_q=\Delta_m$~\cite{Liu2019Magnon,Li2021Strong,Jin2023magnon,Tang2015Quantum}. The transversal coupling strength $J/\gamma=35$, the decay rate $\kappa/\gamma=1$ and the relative phase $\theta=0$, where $\gamma/2\pi=1$ MHz is used as an energy unit throughout the paper.}\label{DqDm}
\end{figure}

Figure~\ref{DqDm} is devoted to confirming Eq.~(\ref{OptDelta1}), for which we employ the optimized condition in Eq.~(\ref{OptOq}) and set $\theta=0$ for simplicity. In the space of $\Delta_q/J$ and $\Delta_m/J$, it is interesting to find that the single-magnon occupation $P_1$ is always maximized with the optimized condition $\Delta_q\Delta_m\approx J^2$ in Eq.~(\ref{OptDelta1}), which is marked by the black dashed line. In particular, we have $P_1\sim0.11$ when $\Delta_m/J=3/4$ and $\Delta_q/J=4/3$. For example, the optimal condition~\cite{Liu2019Magnon,Li2021Strong,Jin2023magnon,Tang2015Quantum} found for the resonant case is confirmed and marked by the red triangle.

\begin{figure}[htbp]
\centering
\includegraphics[width=0.9\linewidth]{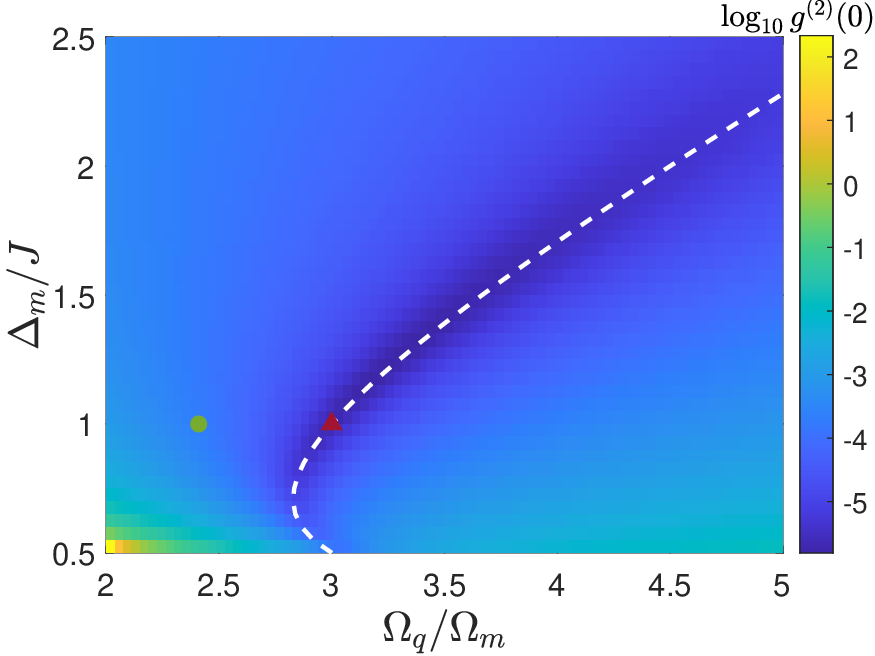}
\caption{Logarithm function of the equal-time second-order correlation function $\log_{10}g^{(2)}(0)$ in the space of the ratios $\Delta_m/J$ and $\Omega_q/\Omega_m$. The white dashed line follows the optimal ratio $\Omega_q/\Omega_m$ in Eq.~(\ref{OptOq}). The red triangle distinguishes the resonant case with $\Delta_q=\Delta_m$~\cite{Liu2019Magnon,Li2021Strong,Jin2023magnon} and the green circle marks the less optimized results obtained in Refs.~\cite{Tang2015Quantum,Xie2020Quantum}. $\Delta_q$ is set by the optimized condition in Eq.~(\ref{OptDelta1}) and the other parameters are set the same as Fig.~\ref{DqDm}.}\label{OqOm}
\end{figure}

In Fig.~\ref{OqOm}, we plot the $g^{(2)}$ function to check Eq.~(\ref{OptOq}) with respect to the ratios $\Delta_m/J$ and $\Omega_q/\Omega_m$. It is found that the optimized point $\Omega_q/\Omega_m$ for the minimization of $g^{(2)}(0)$ varies with respect to $\Delta_m/J$. In particular, the $g^{(2)}$ function is minimized to be $g^{(2)}(0)\sim10^{-5.3}$ when $\Omega_q/\Omega_m=2.85$ for $\Delta_m/J=0.8$, $g^{(2)}(0)\sim10^{-5.7}$ when $\Omega_q/\Omega_m=3$ for $\Delta_m/J=1$, $g^{(2)}(0)\sim10^{-5.8}$ when $\Omega_q/\Omega_m=3.23$ for $\Delta_m/J=1.2$, and $g^{(2)}(0)\sim10^{-5.4}$ when $\Omega_q/\Omega_m=4.5$ for $\Delta_m/J=2$. It is straightforward to verify that these optimized ratios $\Omega_q/\Omega_m$ are exactly the same as the analytical result in Eq.~(\ref{OptOq}), which is marked by the white dashed line. Around the red triangle, our analytical expression in Eq.~(\ref{OptOq}) under the resonant condition confirms the optimal ratio $\Omega_q/\Omega_m=3$ for the blockade numerically obtained in Refs.~\cite{Liu2019Magnon,Li2021Strong,Jin2023magnon}, which is more accurate than that (see the green circle) analytically obtained in Refs.~\cite{Tang2015Quantum,Xie2020Quantum} on an inaccurate assumption $C_{g2}=0$.

\begin{figure}[htbp]
\centering
\includegraphics[width=0.9\linewidth]{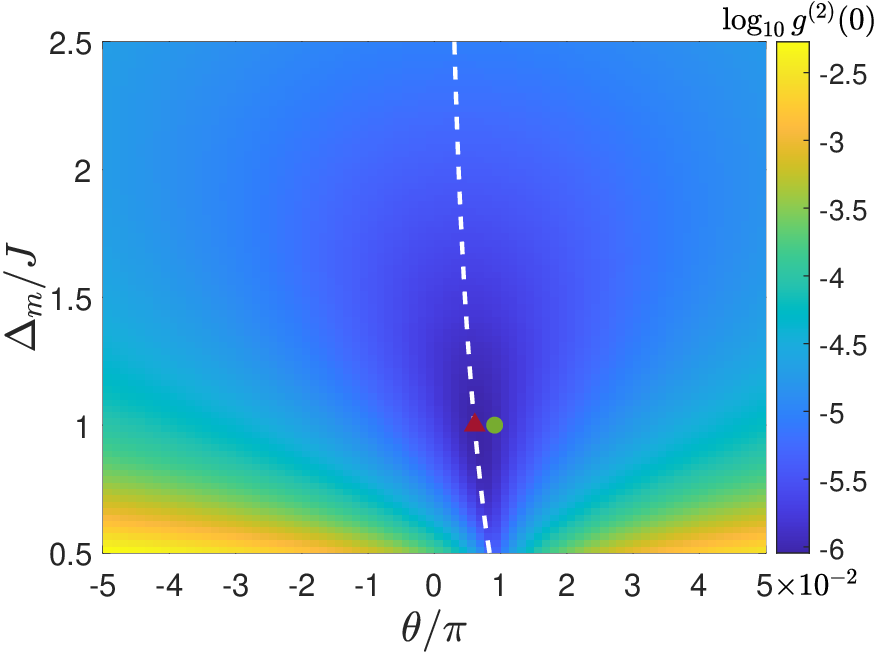}
\caption{Logarithmic function of the equal-time second-order correlation function $\log_{10}g^{(2)}{(0)}$ vs the relative phase $\theta$ and the ratio $\Delta_m/J$. The white dashed line follows the optimal phase $\theta_{\rm opt}$ given in Eq.~(\ref{OptTheta}). The red triangle and the green circle describes our result and the existing one~\cite{Tang2015Quantum,Xie2020Quantum} for the resonant case $\Delta_m=\Delta_q$, respectively. $\Delta_q$ and $\Omega_q$ are set by the optimized conditions in Eqs.~(\ref{OptDelta1}) and (\ref{OptOq}), respectively. The other parameters are set the same as Fig.~\ref{DqDm}.}\label{theta}
\end{figure}

Following Eqs.~(\ref{OptDelta1}) and (\ref{OptOq}), the optimized condition~(\ref{OptTheta}) about the relative phase $\theta$ between the probing and driving fields is confirmed in Fig.~\ref{theta}. It is found that the $g^{(2)}$ function of the stable single-magnon state can be further reduced by an optimal relative phase. In particular, when the driving and probing fields are in phase, i.e., $\theta=0$, the $g^{(2)}$ function is about $g^{(2)}(0)\sim10^{-5.3}$ for $\Delta_m/J=0.8$, $g^{(2)}(0)\sim10^{-5.7}$ for $\Delta_m/J=1$, and $g^{(2)}(0)\sim10^{-5.8}$ for $\Delta_m/J=1.2$. In contrast, the purity of the single-magnon state can be further enhanced to $g^{(2)}(0)\sim10^{-6.0}$ by $\theta/\pi=0.7\times10^{-2}$ for $\Delta_m/J=0.8$, by $\theta/\pi=0.6\times10^{-2}$ for $\Delta_m/J=1$, and by $\theta/\pi=0.5\times10^{-2}$ for $\Delta_m/J=1.2$. All these nonvanishing phases can be analytically obtained by $\theta_{\rm opt}$ in Eq.~(\ref{OptTheta}), which is marked by the white dashed line. The red triangle and the green circle denote our result and the previous optimized result~\cite{Xie2020Quantum,Tang2015Quantum}, respectively, for the resonant case.

\subsection{Approach high-quality single-magnon state}\label{highquality}

\begin{table*}[htbp]
\centering
\caption{Comparison of our proposal and the existing ones for a stable single-quantum state.}\label{table}
\begin{threeparttable}
\begin{tabular}{ccccc}
\hline \hline
\multicolumn{1}{c}{Proposals/systems}&\multicolumn{1}{c}{Experiment/Theory}&\multicolumn{1}{c}{$P_1$}&
\multicolumn{1}{c}{$g^{(2)}(0)$}&
\multicolumn{1}{c}{$\rho_{\rm steady}$}\\
\hline
Our work& Theory &  $\sim0.40$ & $\sim10^{-5}$ & $\approx0.6|0\rangle\langle0|+0.4|1\rangle\langle1|+10^{-6}|2\rangle\langle2|$\\
Quantum dot cavity QED system~\cite{Shijders2018Observation}& Experiment & $\sim0.01$  & $\sim10^{-2}$ &$\approx0.99|0\rangle\langle0|+0.01|1\rangle\langle1|+10^{-6}|2\rangle\langle2|$ \\
Cavity QED system~\cite{Tang2019Strong}& Theory & $\sim0.08$ & $\sim10^{-1.5}$ & $\approx0.92|0\rangle\langle0|+0.08|1\rangle\langle1|+10^{-5}|2\rangle\langle2|$ \\
Photonic crystal cavity system\footnotemark~\cite{Zhou2022Environmentally}& Theory & $\sim0.10$ & $\sim10^{-2}$ & $\approx0.90|0\rangle\langle0|+0.10|1\rangle\langle1|+10^{-4}|2\rangle\langle2|$ \\
Double-cavity system\footnotemark~\cite{Wang2020Photon}& Theory & $\sim0.01$ & $\sim10^{-5}$ & $\approx0.99|0\rangle\langle0|+0.01|1\rangle\langle1|+10^{-9}|2\rangle\langle2|$ \\
Hybrid magnonic system\footnotemark~\cite{Xie2020Quantum}& Theory & $\sim0.05$ & $\sim10^{-2}$ & $\approx0.95|0\rangle\langle0|+0.05|1\rangle\langle1|+10^{-4}|2\rangle\langle2|$ \\
\hline \hline
\end{tabular}
\begin{tablenotes}
\item[a] Photonic crystal cavity system via two-photon absorption process of the environment.
\item[b] Double-cavity system relying on the ultrastrong optomechanical interaction, which is actually challenging for the state-of-art experiment~\cite{Wang2020Photon}.
\item[c] Hybrid magnonic system in which the magnon is coupled to the qubit mediated by a microwave cavity.
\end{tablenotes}
\end{threeparttable}
\end{table*}

Running the master equation (\ref{masterequationGeneral}) with all the optimal conditions in Eqs.~(\ref{OptDelta1}), (\ref{OptOq}), and (\ref{OptTheta}), we are now on the stage approaching a steady and high-quality single-magnon state though parametric optimization over $P_1$ and $g^{(2)}(0)$.

\begin{figure}[htbp]
\centering
\includegraphics[width=0.9\linewidth]{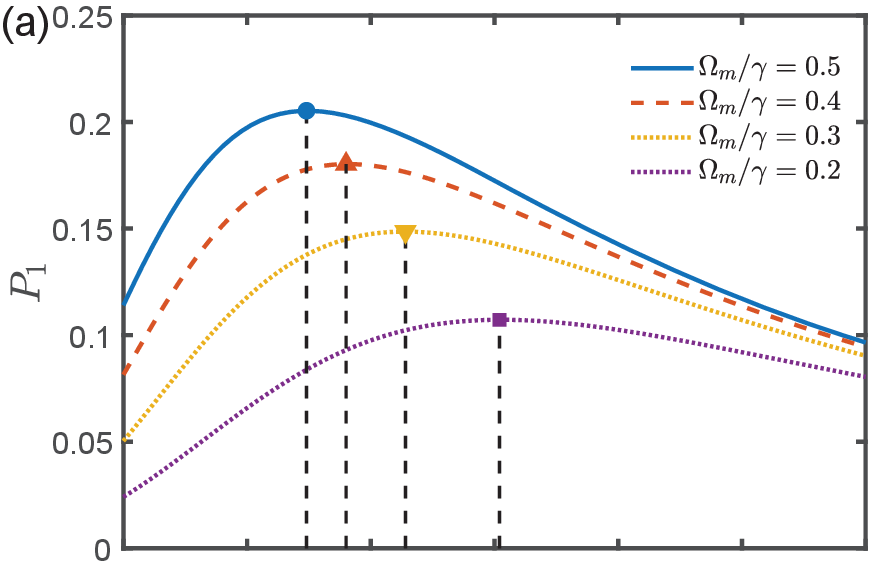}
\includegraphics[width=0.9\linewidth]{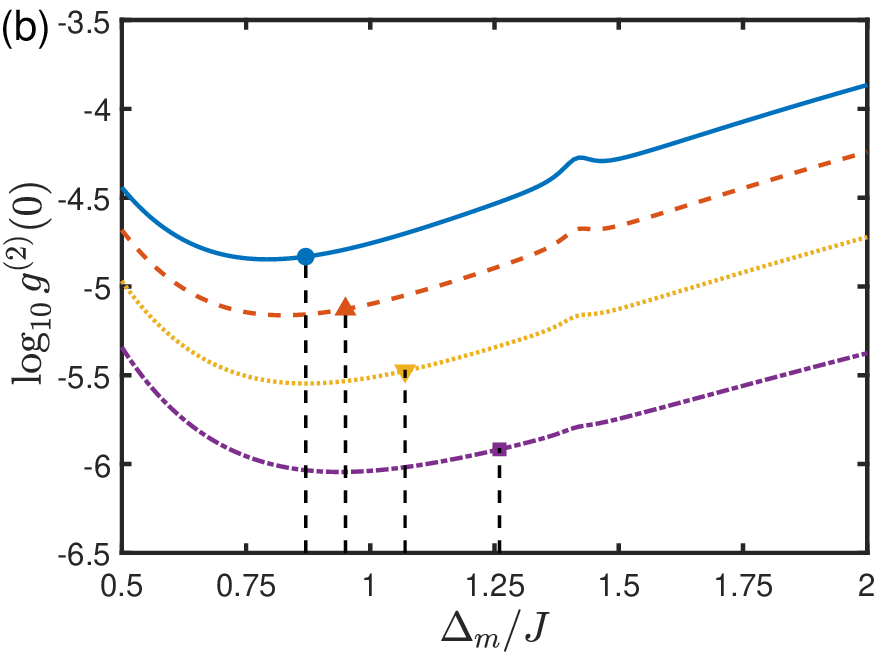}
\caption{(a) Single-magnon probability $P_1$ and (b) logarithmic function of the equal-time second order correlation function $\log_{10}g^{(2)}{(0)}$ versus the ratio $\Delta_m/J$ under various driving intensities $\Omega_m/\gamma$. $\Delta_q$, $\Omega_q$, and $\theta$ are set by the optimized conditions in Eqs.~(\ref{OptDelta1}), (\ref{OptOq}), and (\ref{OptTheta}), respectively. The other parameters are set as $J/\gamma=35$ and $\kappa/\gamma=1$.}\label{OptDm}
\end{figure}

In Fig.~\ref{OptDm} under various driving intensities $\Omega_m/\gamma$, we demonstrate $P_1$ and the $g^{(2)}$ function of the stabilized single-magnon state with respect to the ratio $\Delta_m/J$. Under the constraints in Eqs.~(\ref{OptDelta1}), (\ref{OptOq}) and (\ref{OptTheta}), we are left with three tunable parameters $\Delta_m$, $J$, and $\kappa$ in determining these statistical properties. It is found that a stronger driving intensity $\Omega_m/\gamma$ gives rise to a larger $P_1$ yet accompanied with a larger $g^{(2)}(0)$. In particular, when $\Delta_m/J=1$, i.e., the magnon and qubit are resonant due to $\Delta_q\Delta_m=J^2$, the single-magnon occupation $P_1$ ranges from $0.09$ to $0.20$ and correspondingly, the equal-time second order correlation function $g^{(2)}(0)$ is in the range of $(10^{-6.0}, 10^{-4.8})$ for $\Omega_m/\gamma\in(0.2, 0.5)$. All of these results could be further optimized by tuning the magnon-qubit detuning, i.e., $\Delta_m/J\neq1$. For $\Omega_m/\gamma=0.2$, $P_1$ is maximized to be $P_1\sim0.11$ and $g^{(2)}(0)\sim10^{-5.9}$ when $\Delta_m/J=1.26$; for $\Omega_m/\gamma=0.3$, $P_1\sim0.15$ and $g^{(2)}(0)\sim10^{-5.5}$ when $\Delta_m/J=1.07$; for $\Omega_m/\gamma=0.4$, $P_1\sim0.18$ and $g^{(2)}(0)\sim10^{-5.1}$ when $\Delta_m/J=0.95$; and for $\Omega_m/\gamma=0.5$, $P_1\sim0.20$ and $g^{(2)}(0)\sim10^{-4.8}$ when $\Delta_m/J=0.87$. In each curve of Figs.~\ref{OptDm}(a) and~\ref{OptDm}(b) with the same driving intensity, one can locate a maximal $P_1$ or a minimal $g^{(2)}(0)$. The optimized points do not always share the same detuning $\Delta_m$, which implies that a compromise has to be made to balance the two indicators.

\begin{figure}[htbp]
\centering
\includegraphics[width=0.9\linewidth]{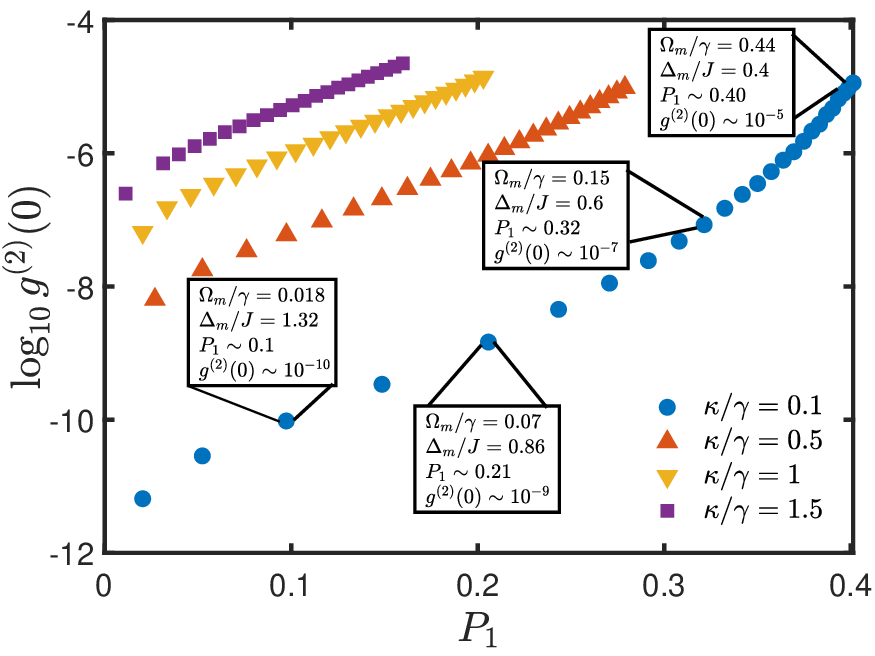}
\caption{Logarithmic function of the equal-time second-order correlation function $\log_{10}g^{(2)}{(0)}$ vs the single-excitation probability $P_1$ for various decay rates $\kappa$ and driving intensity $\Omega_m/\gamma\in(0.005, 0.5)$. Each point is obtained with optimized detuning $\Delta_m/J$. The other parameters are set the same as Fig.~\ref{OptDm}.}\label{P1}
\end{figure}

Due to the probability amplitude $C_{g1}$ in Eq.~(\ref{SteadyAmpa}) that is under the constraint of Eq.~(\ref{OptDelta1}), $P_1$ can be further enhanced by decreasing the decay rate $\kappa$. In Fig.~\ref{P1}, we demonstrate the $g^{(2)}$ function versus $P_1$ with optimized $\Delta_m/J$ for various driving intensities and decay rates. Each point is obtained by locating the optimized $\Delta_m/J$ in a similar way as that in Fig.~\ref{OptDm}(a). One can see that $\log_{10}g^{(2)}{(0)}$ and $P_1$ manifest an almost linear relationship for various $\kappa$ when $\Omega_m/\gamma$ runs from $0.005$ to $0.5$. More important, it is found that when the decay rate is as low as $\kappa/\gamma=0.1$, a single-magnon state presents with a large brightness and a high-degree of purity, e.g., $P_1\sim0.21$ and $g^{(2)}(0)\sim10^{-9}$ when $\Omega_m/\gamma=0.07$ and $\Delta_m/J=0.86$; $P_1\sim0.32$ and $g^{(2)}(0)\sim10^{-7.0}$ when $\Omega_m/\gamma=0.15$ and $\Delta_m/J=0.6$; and $P_1\sim0.40$ and $g^{(2)}(0)\sim10^{-5.0}$ when $\Omega_m/\gamma=0.44$ and $\Delta_m/J=0.4$.

All these results prevail the existing ones in a similar scenario. In the quantum dot-cavity system, the single-photon state with $P_1\sim0.06$ and $g^{(2)}(0)\sim10^{-2}$ has been proposed in theory~\cite{Tang2015Quantum}; and that with $P_1\sim0.01$ and $g^{(2)}(0)\sim10^{-2}$~\cite{Shijders2018Observation} can be experimentally demonstrated. In a cavity QED system~\cite{Tang2019Strong}, the single-photon state can achieve $P_1\sim0.08$ and $g^{(2)}(0)\sim10^{-1.5}$ by using the optical Stark shift, and can achieve $P_1\sim0.01$ and $g^{(2)}(0)\sim10^{-5}$ in a double-cavity system with ultrastrong optomechanical coupling. Using the two-photon absorption process of environment~\cite{Zhou2022Environmentally}, the brightness of the single-photon state can be improved to $P_1\sim0.1$ at the cost of a suppressed purity of $g^{(2)}(0)\sim10^{-2}$. In a hybrid system established by the indirect interaction between magnon and qubit mediated by a microwave cavity~\cite{Xie2020Quantum}, the single-magnon state is found to end up with $P_1\sim0.05$ and $g^{(2)}(0)\sim10^{-2}$ when the second-excitation probability is improperly omitted in the calculation of the optimized conditions. In Tab.~\ref{table}, we list some representative results for $P_1$, $g^{(2)}(0)$, and the steady-state density matrix $\rho_{\rm steady}$ of the interested mode.

\section{General situation in the presence of $N$-magnons}\label{generalmagnon}

Our proposal can be extended to consider the statistical properties of only one of the $N$ magnon modes of the same frequency $\omega_m$ that are uniformly coupled to a common qubit with the coupling strength $J$. In this general situation, the preceding optimized conditions for the ratio of the probing intensity to the driving intensity $\Omega_q/\Omega_m$ and the relative phase between two fields $\theta$ are scale free and only that for the detunings $\Delta_q\Delta_m/J^2$ in Eq.~(\ref{OptDelta1}) is modified to scale with $N$. The total Hamiltonian~(\ref{Hamtot}) including the probing field on qubit and the driving fields on magnons is now written as
\begin{equation}\label{HamtotGeneral}
\begin{aligned}
&H_{\rm tot}=\left[\Omega_m\sum_{j=1}^{N}m^\dagger_je^{-i\omega t}+\Omega_q\sigma_+e^{-i(\omega t+\theta)}+{\rm H.c.}\right]\\
+&\omega_q\sigma_+\sigma_-+\omega_m\sum_{j=1}^{N}m^\dagger_jm_j+J\sum_{j=1}^{N}\left(m_j\sigma_++m^\dagger_j\sigma_-\right),
\end{aligned}
\end{equation}
where $m^\dagger_j$ and $m_j$ are the creation and annihilation operators of the $j$th magnon mode and $\theta$ remains as the relative phase of the fields. Then, in the rotating frame with respect to $H'=\omega(\sum_{j=1}^Nm^\dagger_jm_j+\sigma_+\sigma_-)$, the effective Hamiltonian becomes
\begin{equation}\label{HameffGeneral}
\begin{aligned}
&H_{\rm eff}=\Omega_m\sum_{j=1}^N\left(m^\dagger_j+m_j\right)+\Omega_q\left(\sigma_+e^{-i\theta}+\sigma_-e^{i\theta}\right)\\
+&\Delta_q\sigma_+\sigma_-+\Delta_m\sum_{j=1}^Nm^\dagger_jm_j+J\sum_{j=1}^N\left(m_j\sigma_++m^\dagger_j\sigma_-\right).
\end{aligned}
\end{equation}

Following the analytical model~\cite{Kar2013Single,Tang2015Quantum,Flayac2017Unconventional,
Tang2019Strong,Liu2019Magnon,Xie2020Quantum,Wu2021Phase,Wang2022Hybrid}, the non-Hermitian Hamiltonian in Eq.~(\ref{Hamnon1}) and the approximated system state in Eq.~(\ref{purestate1}) in the low-excitation subspace are modified to be
\begin{equation}\label{HamnonGeneral}
\begin{aligned}
&H_{\rm non}=H_{\rm eff}-i\frac{\kappa}{2}\left(\sigma_+\sigma_-+\sum_{j=1}^Nm_j^\dagger m_j\right),\\
&|\psi\rangle=|\psi_0\rangle+|\psi_1\rangle+|\psi_2\rangle,
\end{aligned}
\end{equation}
respectively. Here $H_{\rm eff}$ is the effective Hamiltonian in Eq.~(\ref{HameffGeneral}) and the subscript in $|\psi_k\rangle$, $k=0,1,2$, implies the state with $k$ excitations. In particular, we have
\begin{equation}\label{Zerostate}
\begin{aligned}
&|\psi_0\rangle=C_{g_0}|g\rangle|0\rangle^{\otimes N},\\
&|\psi_1\rangle=C_{e_0}|e\rangle|0\rangle^{\otimes N}+\sum_{j=1}^NC_{g_j}m_j^\dagger|g\rangle|0\rangle^{\otimes N},\\
&|\psi_2\rangle=\sum_{j=1}^NC_{e_j}m_j^\dagger|e\rangle|0\rangle^{\otimes N}+\sum_{j,k=1}^NC_{g_{jk}}m_j^\dagger m_k^\dagger|g\rangle|0\rangle^{\otimes N},
\end{aligned}
\end{equation}
where $m_j^\dagger|0\rangle^{\otimes N}$ describes that the $j$th magnon is excited and $C_\alpha$ ($\alpha=g_0, e_0, g_j, e_j, g_{jk}$) is the probability amplitude for the relevant base. Due to the system symmetry, one can focus on the first magnon by assuming that the bases of the same number of excitations share the same amplitude, i.e., $C_{g_1}=C_{g_j}$, $2\le j\le N$, and $C_{g_{12}}=C_{g_{1j}}$, $2<j\le N$. Then, the steady-state amplitudes are
\begin{equation}\label{SteadyAmpN}
\begin{aligned}
&C_{g_1}=\frac{J\Omega_qe^{-i\theta}-\Omega_m\tilde{\Delta}_q}{\tilde{\Delta}_q\tilde{\Delta}_m-NJ^2},\\
&C_{g_{11}}\approx\frac{\sqrt{2}(AC_{g_1}-B)}{2\tilde{\Delta}_m(\tilde{\Delta}_q+\tilde{\Delta}_m)-2J^2},\\
&C_{e_1}\approx\frac{-(\Omega_qe^{-i\theta}\tilde{\Delta}_q-NJ\Omega_m)C_{g_1}}
{\tilde{\Delta}_q(\tilde{\Delta}_q+\tilde{\Delta}_m)}+\frac{B}{J(\tilde{\Delta}_q+\tilde{\Delta}_m)},
\end{aligned}
\end{equation}
where the coefficients $A$ becomes $N$-dependent
\begin{equation}\label{GeneralOq}
A=J\Omega_qe^{-i\theta}-\left[(\tilde{\Delta}_q+\tilde{\Delta}_m)+\frac{NJ^2}{\tilde{\Delta}_q}\right]\Omega_m,
\end{equation}
which is consistent with Eq.~(\ref{AB}) and $B$ remains invariant. Subsequently, it is found that the optimized condition for maximizing $P_1$ in Eq.~(\ref{OptDelta1}) is scalable with the system size $N$,
\begin{equation}\label{optimalDeltaN}
\Delta_q\Delta_m\approx\tilde{\Delta}_q\tilde{\Delta}_m=NJ^2,
\end{equation}
and the other two conditions about $\Omega_q/\Omega_m$ in Eq.~(\ref{OptOq}) and $\theta$ in Eq.~(\ref{OptTheta}) are found to be scale free.

\begin{figure}[htbp]
\centering
\includegraphics[width=0.9\linewidth]{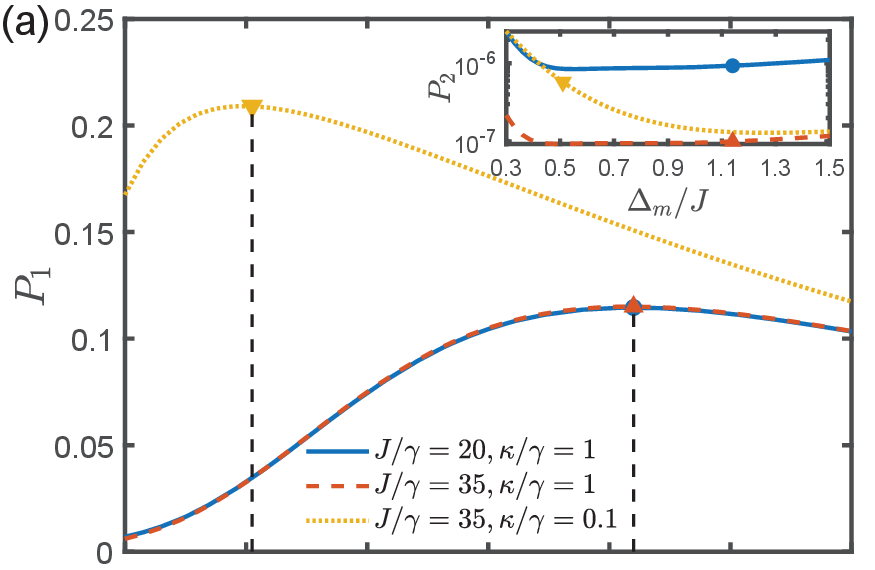}
\includegraphics[width=0.9\linewidth]{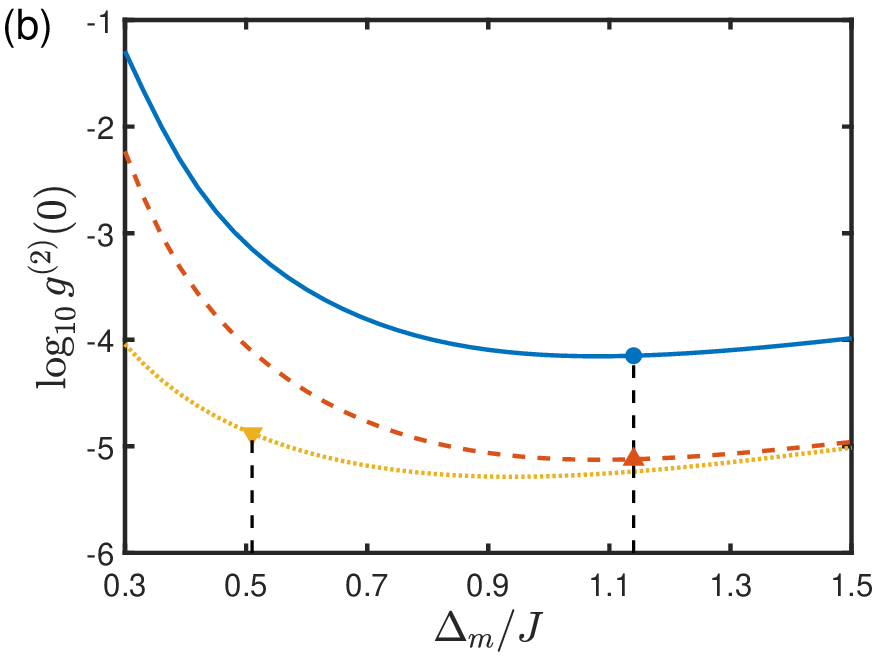}
\caption{(a) Single-magnon probability $P_1$ and (b) Logarithmic function of the equal-time second order correlation function $\log_{10}g^{(2)}{(0)}$ vs the ratio $\Delta_m/J$ under various coupling strengths $J/\gamma$ and decay rates $\kappa/\gamma$. Inset: Double-magnon probability $P_2$ vs $\Delta_m/J$. $\Delta_q$, $\Omega_q$, and $\theta$ are set according to the optimized conditions in Eqs.~(\ref{optimalDeltaN}), (\ref{OptOq}), and (\ref{OptTheta}), respectively. The driving intensity is fixed as $\Omega_m/\gamma=0.44$.}\label{twomodes}
\end{figure}

To numerically simulate the single-magnon state in a double-magnon system, we use the effective Hamiltonian $H_{\rm eff}$ in Eq.~(\ref{HameffGeneral}) and the master equation~(\ref{masterequationGeneral}) with the addition of the decay channel for the second magnon mode. In Figs.~\ref{twomodes}(a) and ~\ref{twomodes}(b) for various coupling strengths $J/\gamma$ and decay rates $\kappa/\gamma$, we demonstrate the single-magnon probability $P_1$ and the $g^{(2)}$ function for the first magnon mode with respect to the ratio $\Delta_m/J$. It is found that roughly a stronger $J/\gamma$ and a weaker $\kappa/\gamma$ yield a larger $P_1$ and a smaller $g^{(2)}(0)$ for any $\Delta_m/J$. In Fig.~\ref{twomodes}, although $P_1$ is not sensitive to the coupling strength in the presence of a weak decay rate, $P_2$ as well as $g^{(2)}(0)$ can be much suppressed by a strong $J/\gamma$. In particular, when $\kappa/\gamma=1$, $P_1$ is optimized as $P_1\sim0.12$ with $\Delta_m/J=1.15$ for both $J/\gamma=20$ and $J/\gamma=35$. In the former case, $P_2\sim10^{-6}$ and $g^{(2)}(0)\sim10^{-4.1}$, while in the latter case, $P_2\sim10^{-7}$ and $g^{(2)}(0)\sim10^{-5.1}$. When the decay rate is as low as $\kappa/\gamma=0.1$, $P_1$ is optimized to be $P_1\sim0.21$ by $\Delta_m/J=0.51$ and $J/\gamma=35$, which is accompanied with $P_2\sim10^{-6}$ and $g^{(2)}(0)\sim10^{-5.0}$.

\section{Conclusion}\label{Conclusion}

In summary, we presents a systematic optimization process to analytically obtain the optimal conditions for the strong magnon blockade, which produces in a long-time limit a high-quality (a large brightness and a high purity) single-magnon state in a hybrid magnon-qubit system by exploiting the strong coupling strength between two components, their weak decay rates, their frequency detunings $\Delta_q$ and $\Delta_m$, the intensity ratio of driving and probing fields $\Omega_q/\Omega_m$, and their relative phase $\theta$. Through parametric optimization, we confirm that under the optimized conditions, i.e., Eqs.~(\ref{OptDelta1}) or (\ref{optimalDeltaN}) for $\Delta_q\Delta_m$, Eq.~(\ref{OptOq}) for $\Omega_q/\Omega_m$, and Eq.~(\ref{OptTheta}) for $\theta$, the single-magnon probability $P_1$ and the equal-time second-order correlation function $g^{(2)}(0)$ could be compromisingly amplified and suppressed, respectively. It is interesting to find that with strong driving intensity and weak decay rates, a single-magnon state with a large brightness and a high purity, i.e., $P_1\sim0.40$ and $g^{(2)}(0)\sim10^{-5}$, can be prepared in a single-magnon system. The optimized conditions presented in our proposal, scaling with the system size or scale free, are actually system independent. Our study thus paves a way toward stabilizing a high-quality single-magnon source for the quantum information process based on a macroscopic spin system, and it is extendable to other bosonic quantum systems.

\section*{Acknowledgments}

We acknowledge grant support from the National Natural Science Foundation of China (Grant No. 11974311).

\bibliographystyle{apsrevlong}
\bibliography{ref2}

\end{document}